\documentclass{aa} 

\usepackage{graphicx}
\usepackage{natbib}
\bibpunct{(}{)}{;}{a}{}{,} 
\usepackage{txfonts}
\begin{document}


 \title{Chemical analysis  of  NGC~6528:  one of the  most  metal-rich bulge globular cluster}

   \author{C. Mu\~{n}oz\inst{1}, D. Geisler\inst{1,5,6}, S. Villanova\inst{1}, I. Saviane\inst{2}, C.C. Cort\'es\inst{1}, B. Dias\inst{2}, R.E. Cohen\inst{3}, F. Mauro\inst{4} \& C. Moni Bidin\inst{4}}


   \institute{Departamento de Astronom\'{i}a, Casilla 160-C, Universidad de
  Concepci\'{o}n, Concepci\'{o}n, Chile.\\
  \email{cesarmunoz@astro-udec.cl}
  \and European Southern Observatory, Casilla 19001, Santiago, Chile.
  \and Space Telescope Science Institute, 3700 San Martin Drive, Baltimore, MD 21218, USA.
 \and Instituto de Astronom\'{i}a, Universidad Cat\'{o}lica del Norte, Av. Angamos 0610, Antofagasta, Chile.
 \and Instituto de Investigación Multidisciplinario en Ciencia y Tecnología, 
Universidad de La Serena. Avenida Raúl Bitrán S/N, La Serena, Chile 
\and  Departamento de Física y Astronomía, Facultad de Ciencias, Universidad de La Serena. Av. Juan Cisternas 1200, La Serena, Chile }
              
\titlerunning{Metal-Rich Bulge GC NGC~6528}
\authorrunning{C. Munoz et al.}


\date{ Accepted}

 
  \abstract
   { The Bulge Globular Clusters (GCs) are  key tracers of this central ancient component of our Galaxy. It is essential to understand their formation and evolution to study that of the bulge, as well as their relationship with the other Galactic GC systems (halo and disk GCs). 
High resolution spectroscopy is a powerful tool for such studies, allowing us to obtain a detailed chemical characterization and kinematics  of the clusters and  to compare their chemical  patterns with those of their halo and disk counterparts.}
   {Our main goals are to obtain  detailed abundances for  a sample of seven red giant members of NGC~6528 in order to characterize its chemical composition and study the relationship of this GC with the bulge, and with other bulge, halo and disk GCs. Moreover, we analyze this cluster's behavior associated with the Multiple Populations (MPs) phenomenon.
}
   {We obtained the stellar parameters and  chemical abundances of light elements (Na, Al), iron-peak elements (V, Cr, Mn, Fe, Co, Ni, Cu), $\alpha$-elements (O, Mg, Si, Ca, Ti) and heavy elements (Zr, Ba, Eu) in seven red giant members of NGC 6528 using high
resolution spectroscopy from FLAMES-UVES.}
   {We obtained in six stars of our sample a mean iron content of [Fe/H]=-0.14$\pm$0.03 dex, in  good  agreement with other studies.   We found no significant internal iron spread. We detected one candidate variable star, which was excluded from the mean  in iron content, we derived a metallicity in this star of [Fe/H]=-0.55$\pm$0.04 dex. Moreover, we found no extended O-Na anticorrelation but instead only an intrinsic Na spread. In addition, NGC~6528  does not  exhibit a  Mg-Al anticorrelation, and no significant spread in  either Mg or Al. The $\alpha$ and iron-peak  elements show good agreement with the bulge field star trend. The heavy elements are slightly dominated by the r-process.
The chemical analysis suggests an origin  and evolution similar to that of typical old Bulge field stars.
Finally, we find remarkable agreement in the chemical patterns of  NGC~6528 and another bulge GC, NGC~6553, suggesting a similar origin and evolution.}
   {}

   \maketitle

\section{Introduction}
The detailed study of each main component of our Galaxy is critical to understand their formation and evolution, as well as that of the Galaxy as a whole. This is graphically illustrated by the many current and new surveys that seek to enlighten our Galactic knowledge, such as: the VVV survey \citep{minniti10}, the Gaia-ESO survey \citep{gilmore12}, SDSS-IV \citep{blanton17}, and The Gaia mission.

One of the main components of our Galaxy is the bulge, which, being likely the oldest Galactic component (e.g. Cescutti et al. 2017), provides us with invaluable information about the origin and subsequent evolution of our Galaxy. One of the key tracers of the bulge is the large number of GCs concentrated there, as first famously realized by Shapley, which we now recognize form an independent GC system from that of the halo \citep{minniti95}. Despite the observational difficulties associated with detailed investigations of such GCs, particularly crowding in this dense central region of the Galaxy and strong and variable reddening, their importance is such that more and more studies are exploring these fundamental objects.

In general, the GCs that have been studied in the more accessible halo and disk have been characterized by inhomogeneities in their light-element content (C, N, O, Na, Mg and/or Al), most prominently appearing as O-Na and Mg-Al anticorrelations. In fact, the O-Na anticorrelation has been found in almost all Galactic GCs\citep{carretta09a}, with   Ruprecht 106 \citep{villanova13} being the most likely exception, and has even been suggested as a defining characteristic of GCs \citep{carretta09a,carretta09b,carretta10b}.

These chemical patterns must be due to the self-enrichment that GCs suffer in the early stages of their formation, allowing the formation of multiple generations of stars within the GC \citep{gratton04}. Suggested `polluters include intermediate mass AGB stars \citep{dantona02,dantona16}, fast rotating massive MS stars \citep{decressin07,krause13} and massive binaries  \citep{demink10,izzard13}.

The spread in iron content is another peculiarity that has been observed in a smaller fraction of GCs, although in some cases this needs to be corroborated with more detailed studies, as in the case of NGC~3201 \citep{simmerer13,munoz13,mucciarelli15}. In general, some  massive GCs in particular display iron variations, most notably Omega Cen \citep{johnson08,marino11a}, M22 \citep{marino11b,dacosta09}, M54 \citep{carretta10a} and Terzan~5  \citep{origlia11}.

The study of Bulge GCs, although lagging behind until now because of observational issues, has seen increasing efforts recently, allowing important improvements in  our previously very scanty knowledge of these key objects. In our first paper on bulge GCs, focused on NGC~6440 \citep{munoz17}, we  found that apparentlRy  the  evolution of some bulge GCs is not so similar to that of the halo GCs as often postulated. Additionally, the bulge GCs show great agreement with the overall chemistry of  bulge field stars \citep{munoz17}. However, the main problems limiting study of bulge GCs remain the  large and variable extinction and high crowding.

In this second paper, we have studied the bulge GC NGC 6528. It is located at (l=$1^\circ, b=-4^\circ$), at a distance of 0.6 Kpc from the Galactic center  with a nominal reddening of E(B-V)= 0.54 (Harris 1996, 2010 edition). It is a rather well-studied bulge GC, lying in Baade's window and very near another bulge GC, NGC 6522. 
NGC~6528 has been studied by several authors with different instruments and techniques.
 \citet{carretta01} studied four Red Horizontal Branch (RHB) stars using high resolution spectrogrpah HIRES  from the Keck telescope.
\citet{origlia05} studied four Red Giant Branch (RGB) stars using the Near Infrared Spectrometer (NIRSPEC) spectrograph at Keck II, and they achieved to measure the chemical abundances for six different elements. Also, \citet{schiavon17}  reported chemical abundances for six elements, which were obtained by Sloan Digital Sky Survey (SDSS)-III. \citet{saviane12}   studied this cluster using the Calcium triplet technique, \citet{lagioia14}   analyzed this cluster using optical photometry from HST, \citet{cohen17} used  near-infrared photometry, \citet{dias15} studied a large sample of GCs among them NGC~6528 with low-resolution spectroscopy and \citet{zoccali04} used high-resolution spectroscopy and they found
overall metallicity Z$\approx$$Z_{\odot}$. It is particularly interesting that \citet{mauro14} found a possible iron spread in this GC based on the low resolution Ca triplet spectra of seven stars obtained by \citet{saviane12}.
We have made a detailed chemical characterization of this GC, measuring the abundance of 18 elements including  light elements, alpha elements, iron-peak and heavy elements, yielding a much more detailed picture of the chemical evolution of this GC than previous available. 

In section 2 we describe the observations and data reduction, in section 3 we explain the methodology we used to calculate atmospheric parameters, errors  and chemical abundances. In section 4 we present our results concerning the variability in one star of our sample and abundances for iron-peak elements, alpha elements, the Na-O anticorrelation, the Mg-Al  relations and heavy elements. Our findings are used  in Sec. 5 to shed light on the origin of NGC~6528. Finally,  in Section 6 our conclusions are presented.

\section{Observations and data reduction}
 
 We observed red giants towards NGC~6528 with the fiber-fed multiobject FLAMES spectrograph mounted at the ESO VLT/UT2 telescope in Cerro Paranal (Chile) (ESO program ID 093.D-0286, PI S. Villanova). We here analyze  the seven stars  observed with the blue and red arms of the high-resolution spectrograph UVES. FLAMES-UVES data have a spectral resolution of about R$\simeq$47000. The data were taken with central wavelength 580\,nm, which covers the wavelength range 476-684\,nm.  Our S/N is about 30  at 650\,nm. 
 
\begin{figure}
    \includegraphics[width=3.9in,height=3.9in]{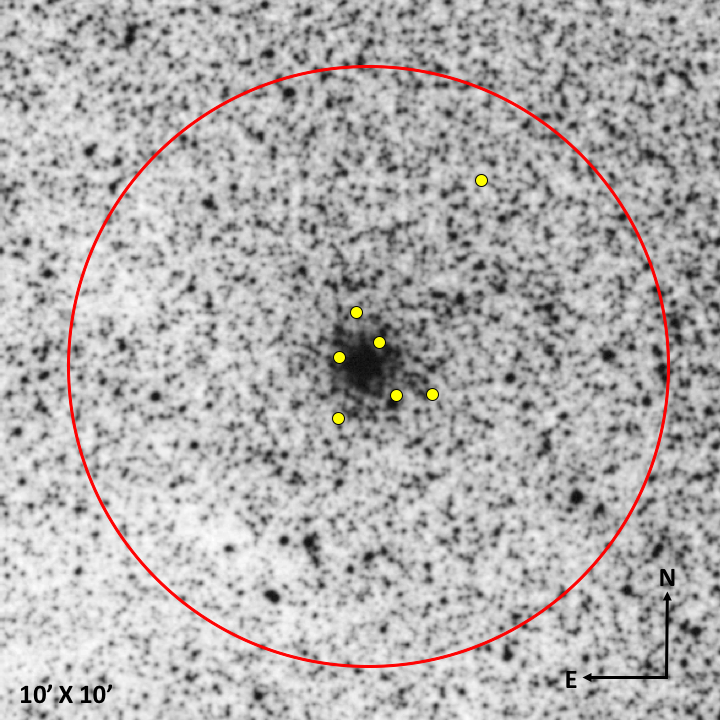}

  \caption{Distribution of  the stars observed  in  NGC~6528 (yellow filled circles). The red circle is the tidal radius (Harris 1996, 2010 edition).}
  \label{spatial}
 \end{figure}

 Six of the seven targets  observed with FLAMES@UVES came from the membership list of NGC~6528 previously published in \citet{saviane12} and \citet{mauro14} using FORS2 and VVV.  Star \#1 is not part of any previous study, but was selected as a likely member based on its position in the CMD. Our subsequent analysis shows that all seven stars are indeed members taking into account the radial velocity, metallicity, CMD position and detailed chemical patterns. Their spatial distribution is shown in Figure \ref{spatial}

\begin{figure}
  \includegraphics[width=3.5in,height=4.1in]{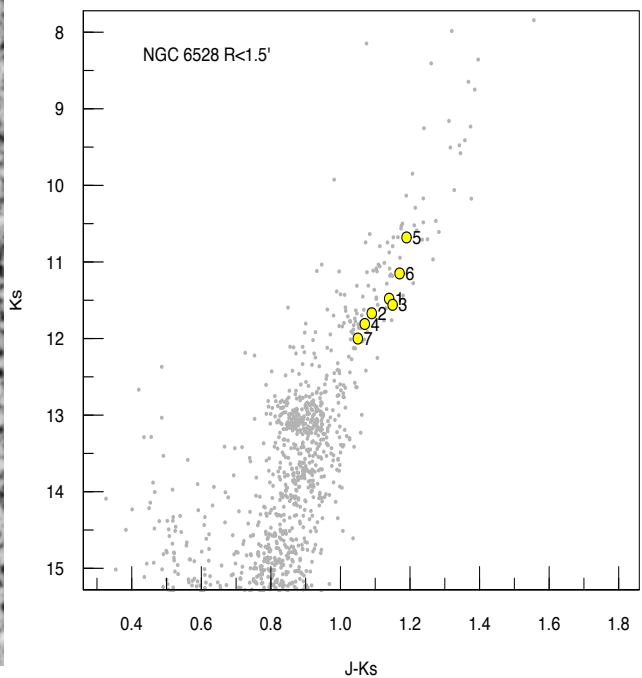}
  \caption{CMD of  NGC~6528 from the VVV survey  corrected  by   the VVV reddening maps \citep{gonzalez12}. The yellow filled circles are our observed UVES sample.}
  \label{CMD}
 \end{figure}

All the stars of our sample  belong to the upper  RGB, as can be clearly seen in the CMD of the cluster (Figure \ref{CMD}). Star  \#5  will be discussed in section  4.1.


Data reduction was performed using the ESO CPL based FLAMES/UVES Pipeline version 5.3.0\footnote{\url{http://www.eso.org/sci/software/pipelines/}} for extracting the individual fiber spectra. Data reduction includes bias subtraction, flat-field correction, wavelength calibration, and spectral rectification.

We subtracted the sky using the Sarith package in IRAF and measured radial velocities using the FXCOR package in IRAF and a synthetic spectrum as a template. The mean heliocentric radial velocity for our seven targets is  210.8 $\pm$ 2.5 km $s^{-1}$. Our velocity dispersion is 6.7 km $s^{-1}$. 

Our mean radial velocity is compatible with the values in the literature: \citet{saviane12} with four stars found a value of 205 $\pm$ 2  km $s^{-1}$, \citet[2010 edition]{harris96}  quotes a value of 206.6 $\pm$ 1.4  km $s^{-1}$,  \citet{carretta01}  found a  value of 210 km $s^{-1}$, \citet{schiavon17}  reported a mean radial velocity of 213 km $s^{-1}$ for two   stars and \citet{origlia05} found a value of 210 
km $s^{-1}$ in a sample of four RGB stars. Table \ref{param} lists the stellar parameters of our sample: ID, the J2000 coordinates (RA and Dec), J, H, K$_{s}$ magnitudes from VVV PSF photometry (de-reddened), calibrated on the system of 2MASS \citep{mauro14,cohen17}, heliocentric radial velocity, Teff, log(g), metallicity and micro-turbulent velocity ($v_{t}$). Moreover, Table \ref{iron-abun} shows the metallicity values from \citet{saviane12}, \citet{mauro14} and \citet{liu17}.

\begin{figure}
  \includegraphics[width=3.7in,height=3.9in]{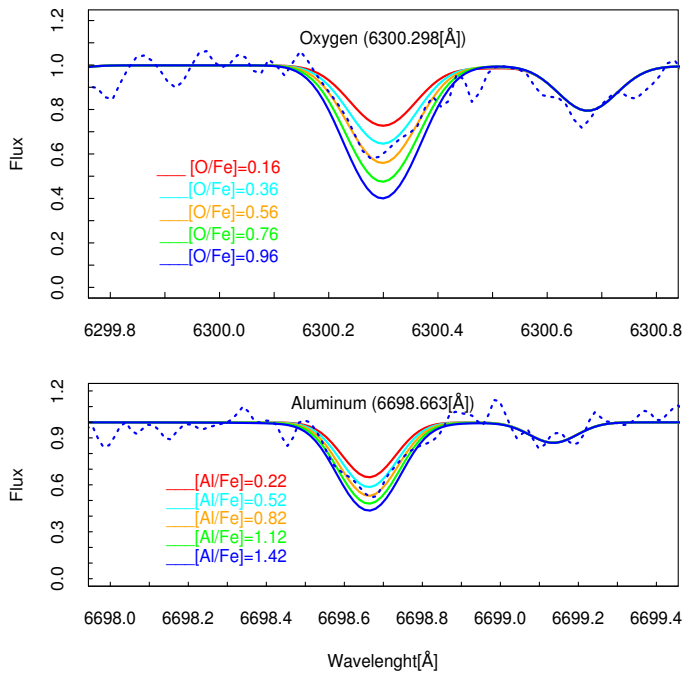}
  \caption{Spectrum synthesis fits for Oxygen (star \#5) and Aluminum (star \#5) lines respectively. The dotted line is the observed spectrum and the solid color lines show the synthesized spectra corresponding to different
abundances.}
  \label{synth}
 \end{figure}

\begin{figure}
  \includegraphics[width=3.5in,height=3.5in]{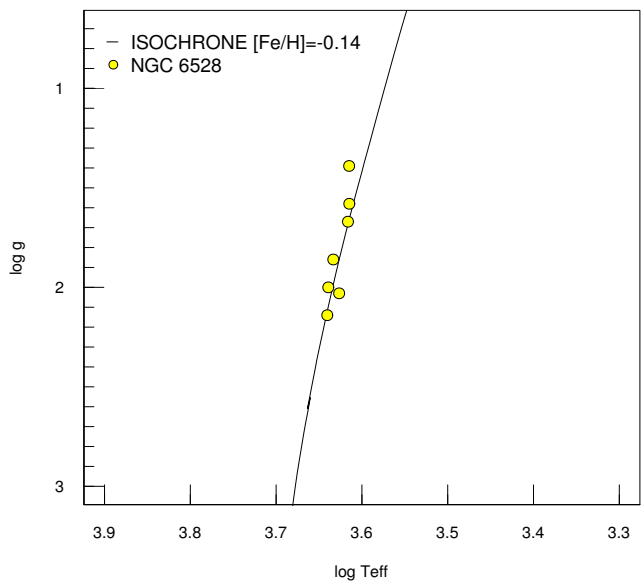}
  \caption{Log g vs log Teff for our sample of seven stars. The overplotted isochrone has a metallicity of -0.14 dex,[$\alpha$/Fe]=+0.20 dex  and age of 11 Gyr \citep{dotter08}.}
  \label{iso}
 \end{figure}


\begin{table*}
\caption{Parameters of the observed stars.}
\label{param} \centering 
\begin{tabular}{ c c c c c c c c c c c c}
\hline 
{\small{}ID} & {\small{}Ra} & {\small{}DEC } & {\small{}J } & {\small{}H } & {\small{}K$_{s}$} & {\small{}RV$_{H}$ } & {\small{}T$_{eff}$} & {\small{}log(g)} & {\small{}{[}Fe/H{]} } & $v_{t}$\tabularnewline

 & {\small{}(h:m:s)} & {\small{}($\,^{\circ}{\rm }$:$^{\prime}$:$^{\prime\prime}$ )} & {\small{}(mag)} & {\small{}(mag)} & {\small{}(mag) } & {\tiny{}(km $s^{-1}$)} & {\small{}{[}K{]} } &  & dex & {\small{}{[}km/s{]} } \tabularnewline
\hline 
1  & 18:04:42.16 &-30:00:49.70  & 12.62  & 11.72  & 11.49 & 198.93$\pm$ 0.21 & 4133 & 1.67  & -0.16 & 1.14    \tabularnewline
2  & 18:04:51.32 & -30:04:05.68 &12.76  & 11.92 & 11.67 &209.91$\pm$0.14   & 4357  & 2.00  & -0.12  & 1.43  \tabularnewline
3  & 18:04:47.76 & -30:03:47.05  &12.71  & 11.82 & 11.56 &215.30 $\pm$0.16 &4232   &2.03   & -0.24  & 1.52  \tabularnewline
4  &18:04:50.13 & -30:02:38.05 &12.88  & 12.02 &11.81 & 216.08$\pm$0.20 & 4369 & 2.14 & -0.11  & 1.42   \tabularnewline
5  &18:04:45.38 & -30:03:46.94 &11.87  &10.95  & 10.68&209.70 $\pm$0.23   & 4118 & 1.58&-0.55  &1.79  \tabularnewline

6  &18:04:51.17  &-30:03:14.95  & 12.32 & 11.43 &11.15 & 218.71$\pm$ 0.19& 4120  & 1.39  &-0.20   & 1.27  \tabularnewline
7  &18:04:48.75  &-30:03:01.88  &13.05  &12.23  & 12.00 &206.91 $\pm$0.18 & 4299  &1.86   & -0.03  & 1.12 \tabularnewline
\hline 
\end{tabular}
\tablefoot{The J,H,Ks magnitudes are de-reddened PSF  photometry of VVV \citep{mauro14,cohen17}.
}

\end{table*}


\begin{table}
\caption{Iron abundances from previous studies.}
\label {iron-abun}
\centering
\small
\begin{tabular}{ c c c c c}

\hline 
\hline
ID. & [Fe/H]$_{this\_work}$ & [Fe/H]$_{S12}$ &{[}Fe/H{]}$_{M14}$&[Fe/H]$_{L17}$\\
 &  &  & & \\
\hline

1	&-0.16$\pm$0.06& - & - & - \\
2	&-0.12$\pm$0.06& - &-0.15 $\pm$0.14 & - \\
3	&-0.24$\pm$0.06& - & -0.27$\pm$ 0.14& -0.07$\pm$0.12 \\
4	&-0.11$\pm$0.06&-0.38 $\pm$ 0.16& -0.50$\pm$0.14 & - \\
5	&-0.55$\pm$0.06& - &-0.17 $\pm$0.14 &-0.15 $\pm$0.09 \\
6	&-0.20$\pm$0.06& -0.22$\pm$0.16 &-0.44 $\pm$0.14 &-0.17$\pm$0.12 \\
7	&-0.03$\pm$0.06& -0.57$\pm$0.16 & -0.64$\pm$ 0.14& -\\

\hline
\end{tabular}
\tablefoot{S12: \citet{saviane12}; M14: \citet{mauro14}; L17: \citet{liu17}\\

}
\end{table}

\section{Atmospheric Parameters and Abundances}
We have analyzed our sample of seven members of  NGC~6528  using  the local thermodynamic equilibrium (LTE) program MOOG \citep{sneden73}.
The procedure which  we used in this paper to calculate the atmospheric parameters is the same that was described in our previous study of NGC 6440 \citep{munoz17}. In  Figure \ref{iso} we show the good agreement between our stellar parameters and an isochrone of similar metallicity as we derive ({[}Fe/H{]}=-0.14 dex) and age of 11 Gyr \citep{dotter08}.

Although NGC 6528 lies towards the  bulge, its location in Baade's Window means the extinction is not as high as for other nearby regions with similar Galactic coordinates. The mean color excess quoted by \citet[2010 edition]{harris96}   is  E(B-V)=0.54 and  \citet{momany03} found a very similar value of  E(B-V)=0.55. Moreover, like the majority of bulge GCs, it is assumed to have differential reddening and certainly high crowding being in Baade's Window, which complicates the  identification of member stars. We determined the stellar parameters directly from the spectra, so our measurement of abundances  is not affected by the effects of reddening.


We used equivalent widths (EWs) of the spectral lines  to obtain the abundances for Ca, Ti, Fe, Co and Ni; the detailed explanation of the method we used to measure the EWs is given in \citet{marino08}. Spectrum synthesis was used to determine the abundances of the other elements (O, Na, Mg, Al, Si, Sc, V, Cr, Mn, Cu, Zr, Ba and Eu), whose lines are affected by blending. We calculated five synthetic spectra of different abundance for each line, and estimated the best-fitting value as the one that minimizes the rms scatter. An example  of this method is showed in Figure \ref{synth} for two lines (Oxygen and Aluminum).  Only lines not contaminated by telluric lines were used. The adopted solar abundances we used are reported in Table \ref{abundances}.



We  performed an internal error analysis  varying $T_{eff}$, log(g), {[}Fe/H{]}, and $v_{t}$ and redetermining abundances of  each  element for star \#3 of our sample, assumed to be representative of the entire sample. 
Parameters were varied by $\Delta T_{eff}$=+41 K, $\Delta$log(g)=+0.14 dex, $\Delta{[}Fe/H{]}$=+0.05 dex, and $\Delta v_{t}$=+0.13 $km$ $s^{-1}$, which we estimated as our typical internal errors. The amount of variation of the parameter was calculated using three stars representative of our sample (\#1,\#3, and \#4)  with  relatively low, intermediate and high effective temperature respectively, according to the procedure that was performed by \citet{marino08}, which we follow in this study. Moreover,  the procedure was used previously in several studies with excellent results \citep{munoz13,munoz17,mura17}.

The error introduced by the uncertainty on  the  EW ($\sigma_{S/N}$) was calculated by dividing the rms scatter  by the square root of the number of the lines used for a given element and a given star. For elements whose abundance was obtained by spectrum-synthesis, the error is given in the output of the fitting procedure.

Finally, the error for each [X/Fe] ratio as a result of  uncertainties in atmospheric parameters and $\sigma_{S/N}$ are listed in Table \ref{error}. The total internal error ($\sigma_{tot}$) is given by:

\begin{center}
\begin{equation}
\sigma_{tot}=\sqrt{\sigma^{2}_{T_{eff}}+\sigma^{2}_{log(g)}+\sigma^{2}_{v_{t}}+\sigma^{2}_{[Fe/H]}+\sigma^{2}_{S/N}}
\end{equation}
\end{center}

In Table \ref{error} we compare the total internal error for each element with the observed error (standard deviation of the sample).

\begin{figure}
  \includegraphics[width=3.8in,height=3.25in]{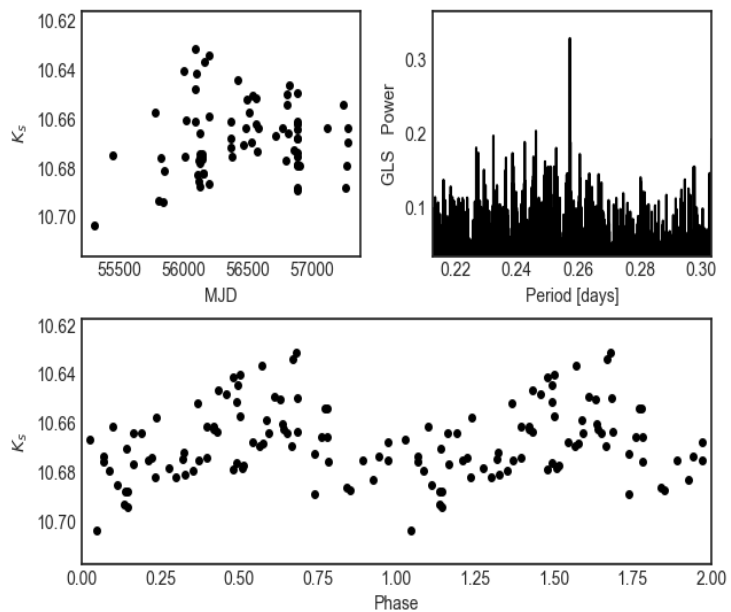}
  \caption{Top left: light curve of the candidate variable star.
Top right: Generalized Lomb Scargle periodogram shows a strong peak at 0.25729 days.
Bottom: Phased light curve of the candidate variable star.}
  \label{variab}
 \end{figure}

 
 
\begin{table*}
\caption{Abundances of the observed stars.}
\label {abundances}
\centering
\begin{tabular}{ l c  c c  c  c c  c  c c }
\hline 
\hline

El. & 1 & 2  & 3 & 4  & 5 & 6 & 7  & Cluster  &Sun  \\
\hline 
$ [$O/Fe$] $   &   0.07  &	0.00  &	0.31  &	0.15&	0.11   &	-0.09   &	-0.09     & +0.07$\pm$0.05   & 8.83 \\
&$\pm$0.03&$\pm$0.04&$\pm$0.04&$\pm$0.06&$\pm$0.05&$\pm$0.03&$\pm$0.05&\\

$ [$Na/Fe$]_{NLTE} $  &	 0.10  &	-0.02 &	0.31  &	0.75 &	0.82  &	0.60     &	0.19  & +0.39$\pm$0.13   & 6.32\\
&$\pm$0.08&$\pm$0.07&$\pm$0.06&$\pm$0.07&$\pm$0.06&$\pm$0.04&$\pm$0.07&\\

$ [$Mg/Fe$] $  &	 0.34  &	0.37   &	0.24   &	0.20  &	0.29   &	0.28   &	0.11     & +0.26$\pm$0.03   & 7.56\\
&$\pm$0.03&$\pm$0.05&$\pm$0.05&$\pm$0.06&$\pm$0.07&$\pm$0.04&$\pm$0.06&\\

$ [$Al/Fe$] $  &	 0.44  &	0.32   &	0.54   &	0.56  &	0.49   &	0.40   &	0.33    & +0.44$\pm$0.04  & 6.43\\
&$\pm$ 0.04&$\pm$0.04&$\pm$0.04&$\pm$0.05&$\pm$0.07&$\pm$0.04&$\pm$0.05  \\

$ [$Si/Fe$] $  &	0.16  &	-0.17   &-0.06   &	0.20  &	-0.10   &-0.07   &	-0.12   & -0.02$\pm$0.05   & 7.61\\
&$\pm$0.08&$\pm$0.06&$\pm$0.06&$\pm$0.07&$\pm$0.06&$\pm$0.04&$\pm$0.08&\\

$ [$Ca/Fe$] $  &	 0.01   &	-0.10   &	0.02   &	0.02 &	0.10   &	0.21   &	-0.05     & +0.03$\pm$0.04   & 6.39\\
&$\pm$0.07&$\pm$0.06&$\pm$0.04&$\pm$0.06&$\pm$0.05&$\pm$0.05&$\pm$0.06&\\

$ [$Sc/Fe$] $  &	 0.32  &	0.22  &	0.38 &	0.41 &	-0.12  &	0.04   &	0.11    & +0.19$\pm$0.07   & 3.12\\
&$\pm$0.07&$\pm$0.06&$\pm$0.06&$\pm$0.07&$\pm$0.07&$\pm$0.04&$\pm$0.07&\\
$ [$Ti/Fe$] $  &0.44  &	0.24   &	0.47  &	0.52 &	0.60   &	0.73     &	0.45  & +0.49$\pm$0.06  & 4.94\\
&$\pm$0.03&$\pm$0.04&$\pm$0.04&$\pm$0.06&$\pm$0.05&$\pm$0.06&$\pm$0.06&\\

$ [$V/Fe$] $   &	0.13&	0.17  &	0.22  &	0.19 &	0.33  &	0.25  &	0.23  & +0.22$\pm$0.02   & 4.00\\
&$\pm$0.06&$\pm$0.06&$\pm$0.06&$\pm$0.06&$\pm$0.05&$\pm$0.04&$\pm$0.06&\\

$ [$Cr/Fe$] $   &	0.07&	-0.07  &	0.31  &	0.10 &	0.18  &	0.26  &	-0.03  & +0.12$\pm$0.05   & 5.63\\
&$\pm$0.10&$\pm$0.10&$\pm$0.10&$\pm$0.11&$\pm$0.06&$\pm$0.06&$\pm$0.08&\\

$ [$Mn/Fe$] $   &	0.40&	0.04 &	0.27  &	0.04 &	0.45  &	0.17 &	0.08  & +0.21$\pm$0.06   & 5.37\\
&$\pm$0.09&$\pm$0.07&$\pm$0.09&$\pm$0.09&$\pm$0.07&$\pm$0.08&$\pm$0.09&\\

$ [$Fe/H$] $   &	 -0.16&	-0.12 &	-0.24  &	-0.11 &	-0.55  &	-0.20  &	-0.03  & -0.20$\pm$0.06   & 7.50 \\
&$\pm$0.03&$\pm$0.02&$\pm$0.02&$\pm$0.03&$\pm$0.04&$\pm$0.03&$\pm$0.03&\\

$ [$Co/Fe$] $  &	 0.74   &	0.52   &	0.64   &	0.61 &	0.54   &	0.83   &	0.53    &+0.63$\pm$0.04  & 4.93\\
&$\pm$0.08&$\pm$0.08&$\pm$0.09&$\pm$0.10&$\pm$0.09&$\pm$0.08&$\pm$0.10&\\

$ [$Ni/Fe$] $  & 0.25  &	0.15  &	0.22  &	0.16 &	0.13 &	0.41  &	0.22  &+0.22$\pm$0.04  & 6.26\\
&$\pm$0.05&$\pm$0.06&$\pm$0.08&$\pm$0.06&$\pm$0.06&$\pm$0.08&$\pm$0.08&\\

$ [$Cu/Fe$] $  & 0.15  &	0.01  &	0.49  &	0.31 &	0.44 &	0.47  &	0.52  &+0.34$\pm$0.07  & 4.19\\
&$\pm$0.09&$\pm$0.11&$\pm$0.12&$\pm$0.13&$\pm$0.10&$\pm$0.10&$\pm$0.12&\\

$ [$Zr/Fe$] $  & -0.19  &-0.22  &-0.14  &-0.17 &	-0.05 &	-0.19  & -0.15  &-0.16$\pm$0.02  & 2.25  \\
&$\pm$0.05&$\pm$0.04&$\pm$0.06&$\pm$0.06&$\pm$0.04&$\pm$0.04&$\pm$0.06&\\

$ [$Ba/Fe$] $  &0.04  &	0.07 & 0.08 &  0.06  &	0.00  &	0.08  & 0.10   & +0.06$\pm$0.01  & 2.34\\
&$\pm$0.06&$\pm$0.05&$\pm$0.06&$\pm$0.07&$\pm$0.04&$\pm$0.04&$\pm$0.06&\\

$ [$Eu/Fe$] $  &	 0.30  &	0.25  &	0.37  &0.31  &	0.13  &	0.16  &	0.26   & +0.25$\pm$0.03 & 0.52\\
&$\pm$0.04&$\pm$0.05&$\pm$0.05&$\pm$0.05&$\pm$0.04&$\pm$0.04&$\pm$0.06&\\

\hline

\end{tabular}
\tablefoot{
Columns 2-8: abundances of the observed stars in NGC 6528. Column 9: mean abundance for the cluster, we include the statistical errors obtained from the mean. Column 10: abundances adopted for the Sun in this paper. Abundances for the Sun are indicated as log$\epsilon$(El.).\\
The errors presented for each abundance, in columns 2-8, was calculated by dividing the rms scatter by the square root of the number of the lines used for a given element and a given star. For elements whose abundance was obtained by spectrum-synthesis, the error is the output of the fitting procedure.

}
\end{table*}


\begin{table*}
\caption{Estimated errors on abundances for representative star of our sample (star \#3), due to errors on atmospherics parameters and to spectral noise, compared with the observed errors.}
\label {error}
\centering
\begin{tabular}{ l c  c c  c  c c  c  c c  c }
\hline 
\hline
	ID   &  $\Delta T_{eff} =41 K $ & $ \Delta log(g)=0.14$  & $\Delta v_{t}= 0.13$	& $ \Delta [Fe/H]=0.05 $ & $\sigma_{S/N}$ & $\sigma_{tot}$  & $\sigma_{obs}$\\		
	\hline					
	$ \Delta ([O/Fe]) $     &	-0.02 &	0.03   &	-0.06   &	-0.04   &	0.04  &	0.09  &	0.14\\ 
	$ \Delta ([Na/Fe]) $  &	0.04 &	-0.03 &	-0.04   &	-0.04   &	0.06 &	0.10 &	0.33\\ 	
	
	$ \Delta ([Mg/Fe]) $    &	0.05 &	0.05   &	0.07   &0.05   &	0.05  &	0.12  &	0.09\\ 	
	$ \Delta ([Al/Fe]) $    &	0.03 &	0.02  &	0.05   &	-0.02   &	0.04  &	0.08  &	0.10\\ 	
	
	$ \Delta ([Si/Fe]) $    &	0.05   &	0.06  &	0.04   &	0.04   &	0.06  &	0.11  &	0.14\\ 
			
	$ \Delta ([Ca/Fe]) $    &	-0.01  &	0.01  &	0.05  &	-0.01  &	0.04  &	0.07 &	0.10\\

	$ \Delta ([Sc/Fe]) $    &	0.03   &	-0.05  &	0.00 &	0.04   &	0.06&	0.09  &	0.19\\ 	
	
	$ \Delta ([Ti/Fe]) $  &	-0.01   &	-0.01 &	0.05   &-0.01  &	0.04&	0.07  &	0.15\\ 	
    
		$ \Delta ([V/Fe]) $    &	0.02  &	0.07  &	0.05  &	 0.06   &	0.06&	0.12  &	0.06\\ 	
     	$ \Delta ([Cr/Fe]) $    &	0.00  &	0.00   &	0.05  &	0.00   &	0.10  &	0.11  &	0.14\\ 
  	  $ \Delta ([Mn/Fe]) $    &	0.00  &	0.03   &	0.06   &	-0.05  &	0.09  &	0.12 &	0.17\\ 	
 
	$ \Delta ([Fe/H]) $     &	0.01   &	0.02   &	0.03   &	0.03   &	0.02  &	0.05  &	0.16\\ 	
	$ \Delta ([Co/Fe]) $    &	0.02   &	-0.03   &	0.06  &	0.02  &	0.09  &	0.12  &	0.12\\ 	
	$ \Delta ([Ni/Fe]) $  &	0.02  &	-0.03   &	0.05   &	0.02   &	0.08  &	0.10 &	0.09\\
    	$ \Delta ([Cu/Fe]) $    &	0.12  &	0.06   &	0.05   &	0.10   &	0.12  &	0.21  &	0.19\\ 
    	$ \Delta ([Zr/Fe]) $    &	 0.11  &	0.07   &	0.06   &	0.08   &	0.17  &	0.12  &	0.05\\ 	
	$ \Delta ([Ba/Fe]) $    &	0.05  &	0.04   &	0.06   &	-0.01   &	0.06  &	0.11 &	0.03\\ 	
	$ \Delta ([Eu/Fe]) $    &	0.01  &	0.06   &	0.03   &	0.03   &	0.05  &	0.09 &	0.08\\

\hline

\end{tabular}
\end{table*}



\section {Results}
Here we   discuss our results in detail, compare them with previous studies of this cluster and also make a general comparison  with other bulge globular clusters analysed up to now.

 \subsection{Variability}
 \label{variability}
 We have checked the variability of all our stars using the VVV survey data \citep{minniti10,saito12}, which  gives us multicolor photometry in five bands: Z (0.87 um),  Y(1.02 um), J (1.25 um), H (1.64 um) and Ks (2.14 um). The survey covers an area of 562 square degrees, with 196 tiles that cover the bulge and 152 covering  the adjacent southern plane. NGC 6528 is located on tile b278 whose central coordinates are RA: 18:04:40.94  DEC: -30:13:18.5 (J2000).
 The  VVV catalogue of aperture photometry for each epoch  was obtained from the Cambridge Astronomical Survey Unit (CASU\footnote{http://apm49.ast.cam.ac.uk/}). The catalogue have position, fluxes and flag which indicate the most probable morphological classification. In this last case,  the flag "-1", which was chosen for this analysis, denote the best-quality photometry of stellar objects, other flags in the catalogue are "0" (noise),  "-2" (borderline stellar), "-7" (sources containing bad pixels), and "-9" (saturated source).

 In order to detect any periodic signals, we computed the Generalized Lomb Scargle (GLS) \citep{zechmesiter09} and Phase Dispersion Minimization (PDM) \citep{stellingwerf78} algorithms of each star using aperture photometry. As a result, we detected significant variability in star \#5.

After this previous confirmation, we verified the variability in this star using PSF photometry. Because  the time series contains multiple points for each epoch, we performed the average for epochs with a few seconds of separation. We obtained a total of 73 epoch in the Ks band from April 2010 to September 2015, in a range of 10.6311 < Ks < 10.7035 mag. We found a period of 0.25729 days, as shown in figure \ref{variab}. The amplitude of the light curve was determined by the Fourier fit, yielding 0.05 mag.

We couldn't identify the type of variable, since it is difficult to classify variability type, due to the few features showed in Ks bands \citep{Catelan13,Alonso2015}.

\begin{figure}[!htbp]
\centering
  \includegraphics[width=3.4in,height=8.6in]{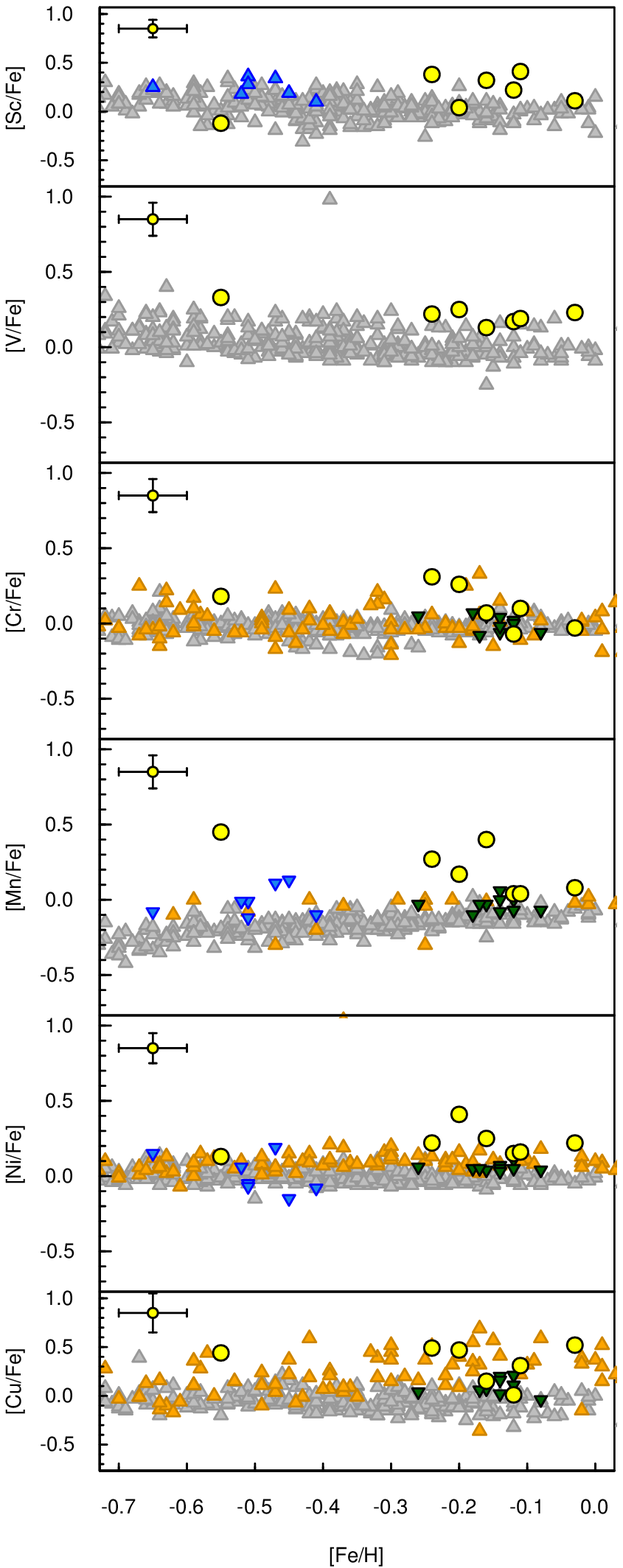}
  \caption{[Sc,V,Cr,Mn,Ni,Cu/Fe] vs [Fe/H]. Filled yellows  circles  are our data for NGC~6528, filled blue triangles: NGC~6440 \citep{munoz17}, filled orange triangles: Bulge field stars \citep{barbuy13,johnson14}, filled gray triangles: halo and disk  stars \citep{fulbright00,francois07,reddy03,reddy06}, filled dark green triangles:  NGC~6553 \citep{johnson14}.}
  \label{iron-ele}
 \end{figure}


 \subsection{Iron}

We found a mean [Fe/H] value for all our sample of seven stars   of  [Fe/H]=$-0.20\pm0.06$ dex. However, the scatter observed  is $\sigma_{obs}$=0.16, which  is  significantly larger than the total expected spread of only 0.05 dex and is thus not due to errors in the  atmospheric  parameters, which are small  in comparison with the scatter (see table \ref{error}). However, it is important to note that our sample is small, only seven stars, and thus small sample statistics could be at fault. Moreover, one of the stars shows a very extreme metallicity compared to the others (star \#5, see table \ref{param}). This star  is  indeed the variable candidate (see Figure \ref{variab} and section \ref{variability}). It is quite possibile that the variability has affected the measurement of the iron abundance of this star in some way, causing a large offset with respect to the cluster mean. Another detailed spectroscopic analysis of this star 
is needed to investigate possible variability effects on the metallicity derivation. We note that,  according to its radial velocity, position in the CMD and atmospheric parameters,  this star is very likely a cluster member.

Nonetheless,  if we remove star \#5 from the sample, assuming that it is an outlier (presumably due to the variability), we obtain a metallicity average of  [Fe/H]=$-0.14\pm0.03$ dex and a scatter of 0.07, essentially what is expected from errors alone. Therefore, we need a larger sample to conclusively prove or disprove an intrinsic metallicity spread. We note that \citet{mauro14} found a mean of  [Fe/H]=-0.18$\pm$0.14 dex with a  spread of 0.20 dex
using low resolution CaT spectra together with NIR photometry for a similar sample of stars (see table \ref{iron-abun}), and suggested a possible intrinsic variation. However, their uncertainties are larger than ours. Note that they found star \#5 to have a metallicity virtually identical with their mean value.

Comparing our results with  \citet{saviane12}  we note that we  have three  stars in common (see table \ref{iron-abun}).  Star  \#7 shows  a significant difference between the two studies while the other two stars do not.  If we use the  metallicity scale of \cite{dias16a,dias16b} for NGC~6528 the average is [Fe/H]=-0.13$\pm$0.07, in excellent agreement with our results.

 \citet{schiavon17} found a metallicity of -0.19 dex in two stars. \citet{origlia05} measured the abundance of four stars using near infrared spectroscopy and found a mean metallicity  of -0.17 dex.  \citet{zoccali04} derived a value of  -0.10 dex  in three stars and \citet[2010 edition]{harris96} quotes a metallicity of -0.11 dex for NGC 6528. \citet{liu17}  found a metallicity of 0.04 dex with a scatter of 0.07 dex, their  S/N is similar to ours. Moreover, we have three stars in common (see table \ref{iron-abun}), with good agreement for star \#3 and \#6. Again, their value for star \#5 is not unusual compared to their other two stars.  And \citet{carretta01} mesaured the metallicity in four RHB stars in NGC~6528, they found a  metallicity of 0.07 dex with  the total errors of $\sim$0.1 dex. In general, we conclude that our mean metallicity is in good  agreement with previous results. Even if [Fe/H] for star \#5 is lower than the mean, its [El./Fe] ratios agrees with the rest of the sample.


\subsection{Iron-peak elements}

The iron peak elements, although they have not been widely studied in bulge GCs, are very useful tools to better understand better their  chemical evolution.    They are produced mainly by SNe and massive stars, the most important polluters of the ISM. \\
In this study, we have measured the abundance of six iron-peak elements: Fe, Sc, V, Cr, Mn and Ni (see table \ref{abundances} and figure \ref{iron-ele}). We here analyze the other elements besides Fe.

In figure \ref{iron-ele} we plot iron-peak element abundance versus [Fe/H]  comparing with two Bulge GCs: NGC~6553 \citep{tang17} and with our previous study of  NGC~6440 \citep{munoz17}. We find for NGC~6528 a super-solar abundance for all the iron peak elements, although slightly  more moderate for the case of V and Ni.  In general, NGC~6528  shows good agreement with  NGC~6553 \citep{tang17} and with bulge field stars from \citet{barbuy13} and \citet{johnson14}.

Although we do not have bulge field star measurements for Sc or V,  we notice that NGC~6528 follow the Galactic disk  field star trend but  at higher abundance, which is typical behavior of bulge stars.

 We found an overabundance in [Mn/Fe]=0.21$\pm$0.06 dex for  NGC~6528 (see Figure \ref{iron-ele}). This overabundance of Mn is an indication of a  formation site rich in Mn, which is mainly produced in  SNeIa \citep{iwamoto99,kobayashi06,cescutti08}.
 Nickel shows in general very good agreement with the bulge field and with NGC~6553.
 
The nucleosynthetic origin of copper is not very clear \citep{mcwilliam05}. Some authors suggest  SNeIa do not contribute significantly to Cu synthesis \citep{shetrone03}. Nevertheless, \citet{matteucci93}  and \citet{mishenina02}    indicate that SNeIa could be an important producer of copper.
In the case of NGC~6528, we found a very good agreement with NGC~6553 and bulge field stars (See figure \ref{iron-ele}).

In summary, NGC~6528 shows good agreement with NGC~6553 for each element compared (Cr, Ni and Cu) and good agreement with bulge field stars  as well.  The super-solar value found for all iron-peak elements  measured in this study could be an indication of a strong early pollution by SNe explosions, mainly  SNeIa. However, the detailed origins of many of these elements remain unclear.

 \citet{carretta01} derivated the abundances for the iron-peak elements (Sc, V, Cr, Mn, and Ni) in four RHB stars. Comparing it with our result, we found a significative difference. Although taking into account the errors in both studies, we can found compatibility between  Sc, Cr, and Ni (see Figure \ref{comp}).

Moreover, our finding show good compatibility with  the result of \citet{barbuy16}, they concluded that abundances of HP1 agree with enrichment by SNII at early epoch, our result show a good agreement with enrichment by SNIa, which happen later. These results are indication that NGC6528 is younger that HP1, which would be expected by its metallicity. In fact, the age measurement by  \citet{ortolani11}  for HP1 is $\sim$13 gyr and \citet{lagioia14} found a age of $\sim$11 gyr  for NGC~6528.

\subsection{$\alpha$ elements}
\label{alpha}
\begin{figure}
\centering
  \includegraphics[width=3.5in,height=8.5in]{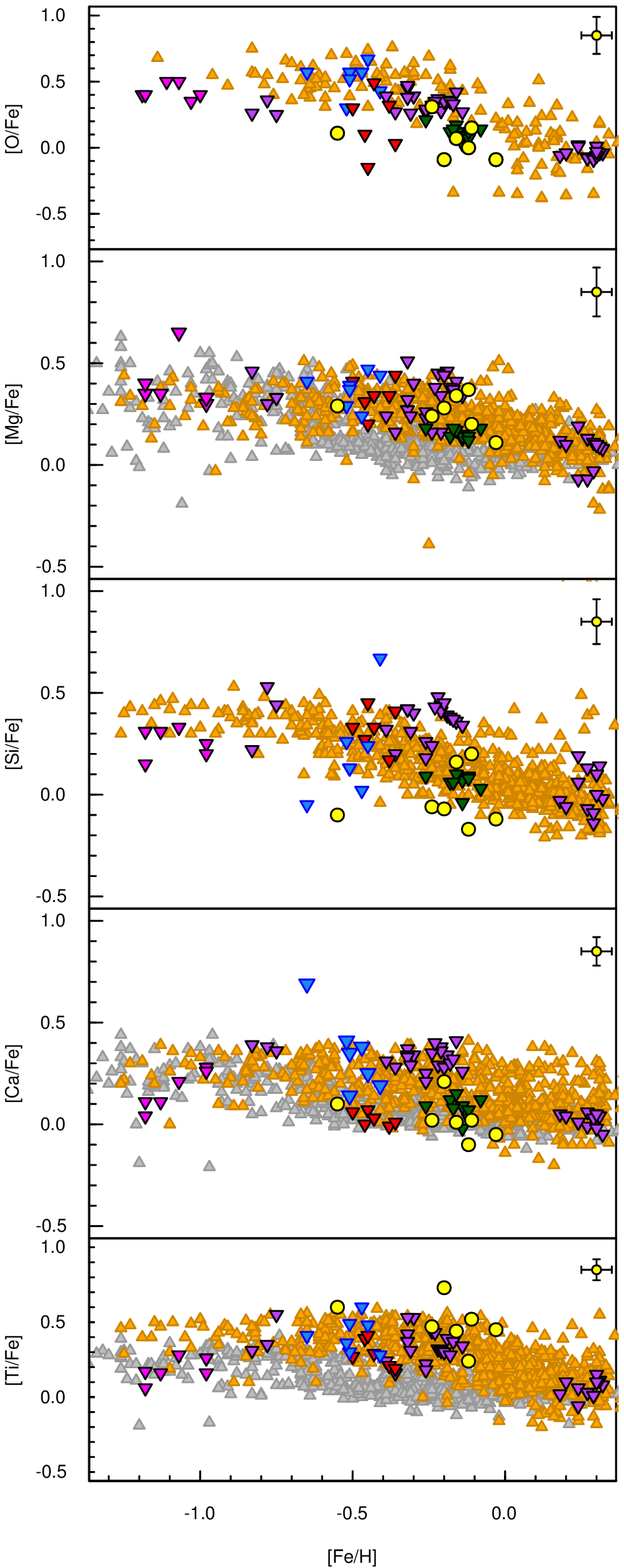}
  \caption{$[$O/Fe$]$,[Mg/Fe], [Si/Fe], [Ca/Fe], [Ti/Fe] vs [Fe/H].  Filled yellow circles are our data for NGC~6528, filled blue triangles: NGC~6440 \citep{munoz17}, filled red triangles:  NGC~6441 \citep{gratton06,gratton07}, filled purple triangles: Terzan 5 \citep{origlia11,origlia13}, filled dark green triangles: NGC~6553 \citep{tang17}, filled magenta triangles: HP1 \citep{barbuy16}, filled orange triangles: bulge field stars\citep{gonzalez12},  filled grey triangles: Halo and Disk fields stars \citep{venn04}.}
  \label{alpha2}
 \end{figure}

We measured five $\alpha$ elements (O, Mg, Si, Ca and Ti). We found that the abundances for O, Si and Ca listed in Table \ref{abundances} (see figure \ref{alpha2}) are  essentially solar. However, Mg and Ti are definitively enhanced (see table \ref{abundances} and Figure \ref{alpha2}). We use  Mg, Si, Ca, and Ti to estimate a mean $\alpha$ abundance. We obtain:

\begin {center}
[$\alpha$/Fe]=$0.19\pm0.03$
\vspace{0.3cm}
\end {center}

Origilia et al (2005) using NIR spectroscopy for  four members of NGC~6528 found a value for [$\alpha$/Fe] $\simeq$ +0.33 $\pm$ 0.01, higher than our value. They obtain the  $\alpha$ abundances using  the same elements used in this study (Mg, Si, Ca and Ti). \citet{carretta01}  measured the alpha abundances in their sample of RHB star, and their values are higher than our, in particular in the case of Titanium, the other elements (Mg, Si, and Ca) are compatible considering the errors.  
Also, \citet{zoccali04} found  a value [$\alpha$/Fe] $\simeq$ +0.1, consistent  with our results (see Figure \ref{comp}).

It is interesting to note that  NGC~6528 shows good agreement with the Bulge field stars with regard to $\alpha$ elements.  In  the case of O, Si and Ca, the cluster follows the trend for the lower part of the field star distribution, while Mg is in the middle of this distribution and Ti in the upper envelope. This certainly is indicative of a similar origin (see section 5).

Notably, the typical $\alpha$ elements that generally show a substantial spread among GCs, O and Mg, do not show evidence for an intrinsic variation in NGC~6528, very reminiscent of the behavior we found for the bulge GC NGC~6553.

\begin{figure}
\centering
  \includegraphics[width=3.5in,height=3.5in]{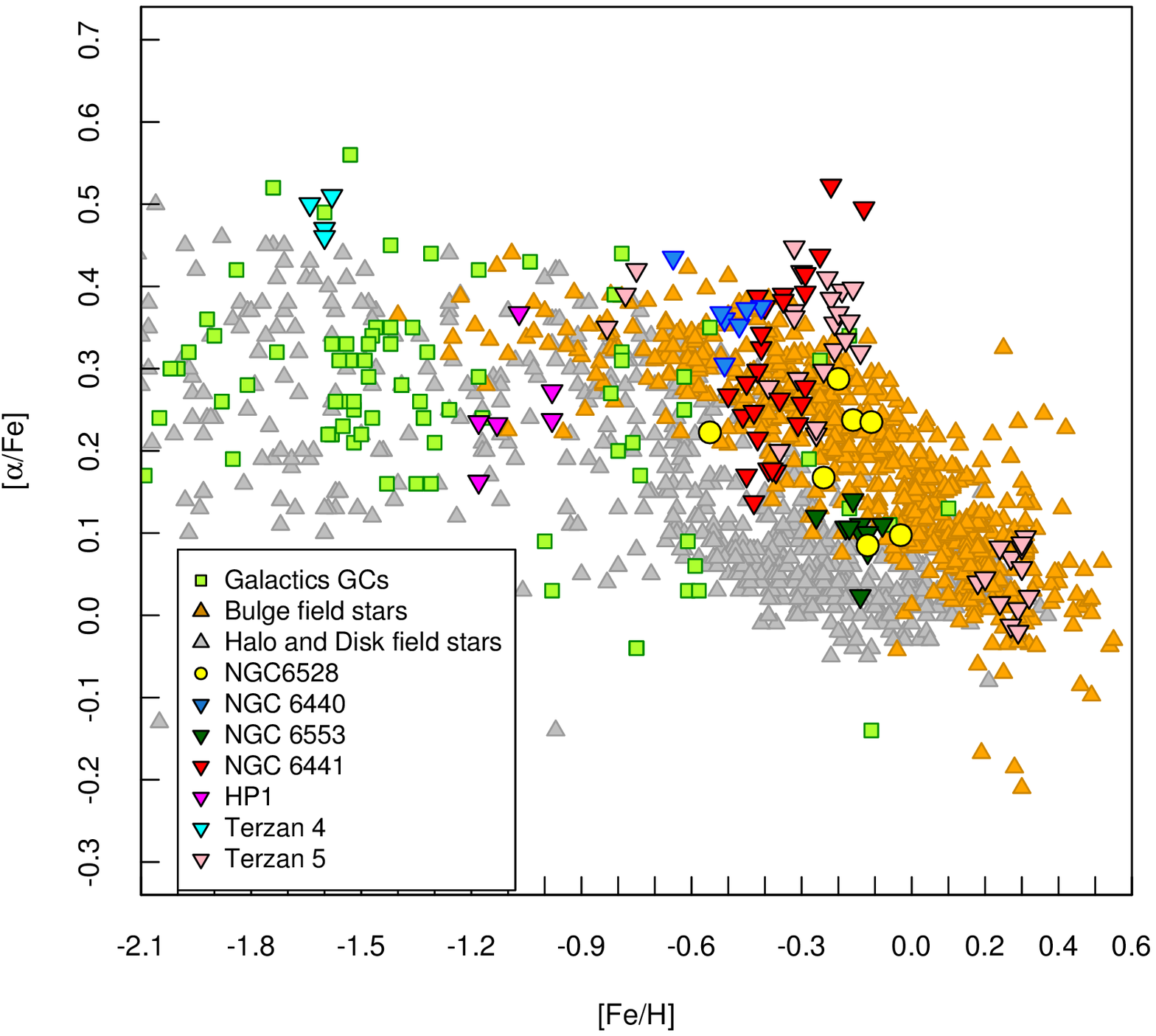}
  \caption{[alpha/Fe] vs [Fe/H]. Filled yellow circles are our data for NGC~6528, filled blue triangles: NGC~6440 (Munoz et al. 2017), filled red triangles: NGC~6441 \citep{gratton06,gratton07}, filled pink triangles: Terzan 5 \citep{origlia11,origlia13}, filled cyan triangles: Terzan 4\citep{origlia04}, filled dark green triangles: NGC~6553 \citep{tang17}, filled magenta triangles: HP1 \citep{barbuy16}, filled orange triangles: bulge field stars\citep{gonzalez12}, filled green square: GCs from \citet{pritzl05},  filled grey triangles: Halo and Disk fields stars \citep{venn04}.}
  \label{alpha1}
 \end{figure}

In general, we see that all the metal rich bulge GCs that we include in this study (NGC~6440, NGC~6441, Terzan~5, NGC~6553 and NGC~6528) show good agreement with the tendency of the bulge field stars (See Figures \ref{alpha2} and  \ref{alpha1}).

As mentioned in our work on  NGC~6440 \citep{munoz17},  the alpha elements in NGC~6528 also show a peculiar behavior.  In the graph of [$\alpha$/Fe] versus [Fe/H], 
NGC~6528 is located at a higher metallicity than the so-called knee, which marks the change between a high rate of pollution of the $\alpha$-elements from SNeII to a pollution coming mainly from SNeIa with less production of  alpha-elements. The patterns, which are observed in O, Si and Ca,  would indicate that NGC~6528 was formed in an area mainly polluted by SNeIa. However, Mg and Ti remain at SNeII-enhanced levels despite the 
high metallicity of this GC.   This argument is supported by \citet{zoccali04} for  O, Ca and Mg, also by \citet{carretta01} for O and Mg,  and by \citet{schiavon17} for the case of Mg. On the other hand,  the argument is not  well supported by \citet{origlia05} (see Figure \ref{comp}). It is necessary take into account that for all  the studies, including ours, the samples are small to be conclusive about it.



\subsection{Na-O anticorrelation}

This is the most famous anticorrelation among GCs, the prime evidence associated with MPs in GCs \citep{carretta09a,carretta09b,gratton12}. However, note that at least one GC, Ruprecht 106, apparently does not show this anticorrelation, although it is likely an extragalactic GC \citep{villanova13}. According to the models \citep{ventura06}, the GCs that show this anticorrelation form a first generation of stars followed by a second generation which is much more Na-rich and O-poor. The nature of the polluters and the details of the pollution process are currently hotly debated.
We did not find an extended O-Na anticorrelation, especially in the case of Oxygen. Comparing the scatter in our measurement, we found a significant intrinsic spread in Na ($\sigma_{obs}$=0.33) while in the case of the Oxygen  ($\sigma_{obs}$=0.14) is very similar to the expected observational errors  ($\sigma_{tot}$=0.09). This behavior is strange because commonly the  Galactic GCs show a large spread in both Oxygen and Sodium.

 \citet{origlia05}  obtained the abundance of Oxygen for four giants, and the average is [O/Fe]=0.31 dex without significative dispersion. In spite of their value is higher than our, are consistent take into account the errors of both (see Figure \ref{comp}).
\citet{carretta01} found a mean of Oxygen of [O/Fe]=0.07 dex in three stars, in the case of Na, Carretta found a mean of [Na/Fe]=0.40 dex in four stars, their values are in excellent agreement with our findings (see Figure \ref{comp}). \citet{zoccali04}  measured the abundance of Oxygen for three RGB stars in NGC~6528. They found an average of [O/Fe]=0.15 dex  with very low dispersion. In addition, they found a mean for [Na/Fe]=0.43 dex with a large dispersion. Although these results are only for three stars, their behavior is consistent with our results  (see Figure \ref{comp}). Finally, \citet{schiavon17} report the chemical abundances for Na in two stars with a value of 0.23 dex and 0.61 dex, despite that are only two stars reported,  these are in concordance with our measurement for Na (see Figure \ref{comp}).

\begin{figure}
\centering
  \includegraphics[width=3.5in,height=3.5in]{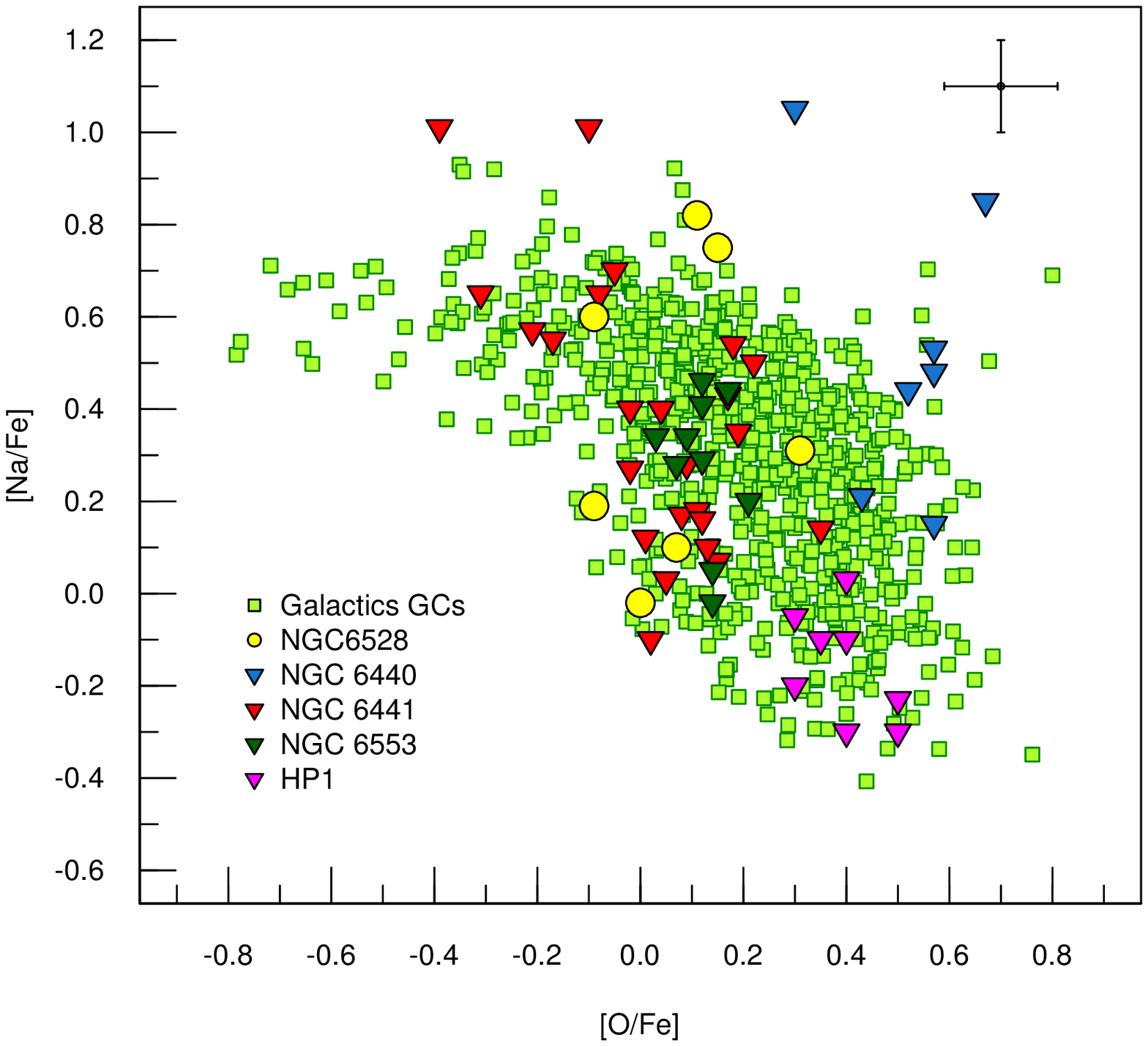}
  \caption{[O/Fe] vs [Na/Fe].
  Filled yellow circles are our data for NGC~6528, filled blue triangles: NGC~6440 \citep{munoz17}, filled red triangles: NGC6441 \citep{gratton06,gratton07}, filled dark green triangles: NGC~6553 \citep{tang17}, filled magenta triangles: HP1 \citep{barbuy16},  filled green square: Galactic GCs from \citet{carretta09b}.}
  \label{o-na}
 \end{figure}

When we study the Na-O anti-correlation  for additional bulge GCs (see figure \ref{o-na}), we note that NGC~6528 (this study), NGC 6440, NGC 6553 and HP1 show a  vertical O-Na trend, marked by a significant spread in Na, but not in Oxygen. In fact,  the bulge GCs show a low intrinsic oxygen scatter, consistent with no true spread. We must bear in mind that the samples are in general small. Therefore, we need to check this behavior with a larger sample. If this trend holds, it would highlight a chemical evolution difference in the bulge GCs compared to typical halo GCs.


\subsection{Mg-Al}

 \citet{carretta09b}   found a Mg-Al anticorrelation in a large sample of Galactic GCs. The Mg-Al chain plays an important role in this correlation,  However, very high temperatures are necessary to achieve it. Aluminum  shows a large spread in many Galactic GCs, particularly those with a metallicity lower than [Fe/H] = - 1.1 dex \citep{meszaros15}. Also, \citet{pancino17} verify that the Mg-Al anticorrelation extension depends on cluster mass and metallicity.
 
 \citet{carretta01} measured the abundances of Magnesium in three RHB stars, the average obtained is [Mg/Fe]=0.14 dex. \citet{origlia05} obtained the abundances of Magnesium in four giants stars, the average obtained is [Mg/Fe]=0.35 dex. \citet{zoccali04} found a mean in Magnesium of [Mg/Fe]=0.07 dex, in three stars. Finally, \citet{schiavon17} reported the chemical abundances for Magnesium and Aluminum in two stars, the averages found by them are: [Mg/Fe]= 0.15 dex and [Al/Fe]=0.28 dex. These studies have small samples or in other cases have measurement only for  Magnesium,  in this context our results are compatible with these studies (see Figure \ref{comp}).

\begin{figure}
\centering
  \includegraphics[width=3.5in,height=3.5in]{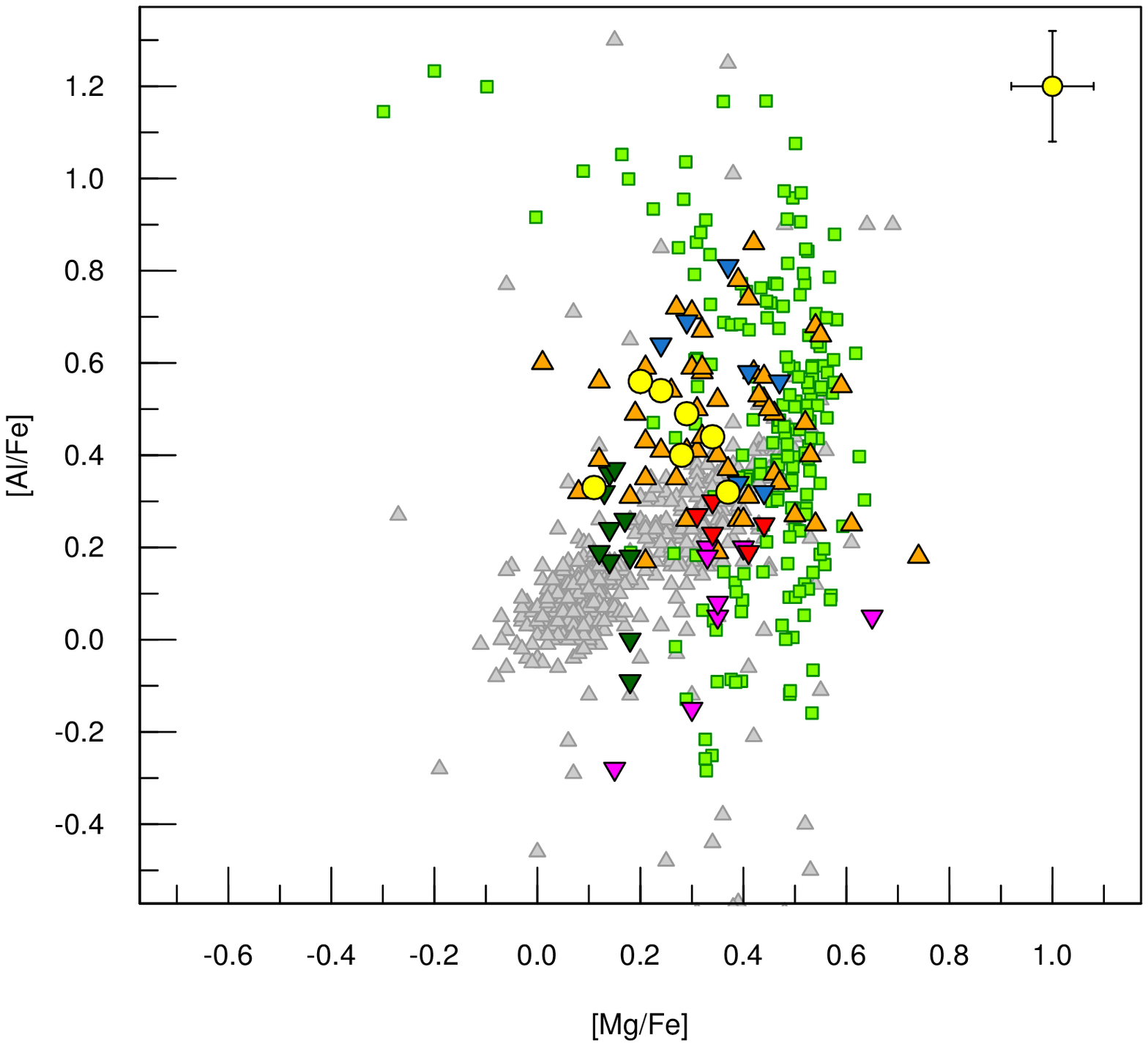}
  \caption{[Mg/Fe] vs [Al/Fe]. Filled yellow  circles are our data for NGC~6528, filled blue triangles: NGC~6440 (Munoz et al. 2017), filled red triangles:  NGC6441 \citep{gratton06},  Filled Yellow triangles: Terzan 5 \citep{origlia11}, filled dark green triangles: NGC~6553 \citep{tang17}, filled magenta triangles: HP1 \citep{barbuy16}, filled orange triangles: bulge field stars \citep{lecureur07}, filled green square: GCs from \citep{carretta09b},  filled grey triangles: Halo and Disk fields stars \citep{fulbright00,reddy03,reddy06,barklem05,cayrel04}.}
  \label{mg-al-na}
 \end{figure}

We find no Mg-Al anticorrelation in NGC 6528 (see Figure \ref{mg-al-na}). Indeed, we find no significant spread  in Al ($\sigma_{obs}$=0.10) or in Mg ($\sigma_{obs}$=0.09), similar to the case for NGC~6553 and  NGC~5927 \citep{mura17}, another metal-rich GC from the disk, but in contrast with NGC 6440, which does show a Mg-Al anticorrelation.
From our study \citep{munoz17}, this bulge globular cluster does not show a large extension in the Mg-Al anticorrelation, but it has a significant spread in Al ($\sigma_{obs}$=0.18), unlike NGC~6528.

\subsection{Neutron-Capture elements}
Analyzing these elements helps us to understand better the processes involved in the creation of MPs in GCs. E.g., low-mass AGB stars  produce mainly s-elements  like Ba (Gallino et al. 1998; Straniero et al. 2006),  and SNeII explosions  are associated mainly with  r-process element production.

\begin{figure}[htb]
  \includegraphics[width=3.5in,height=7in]{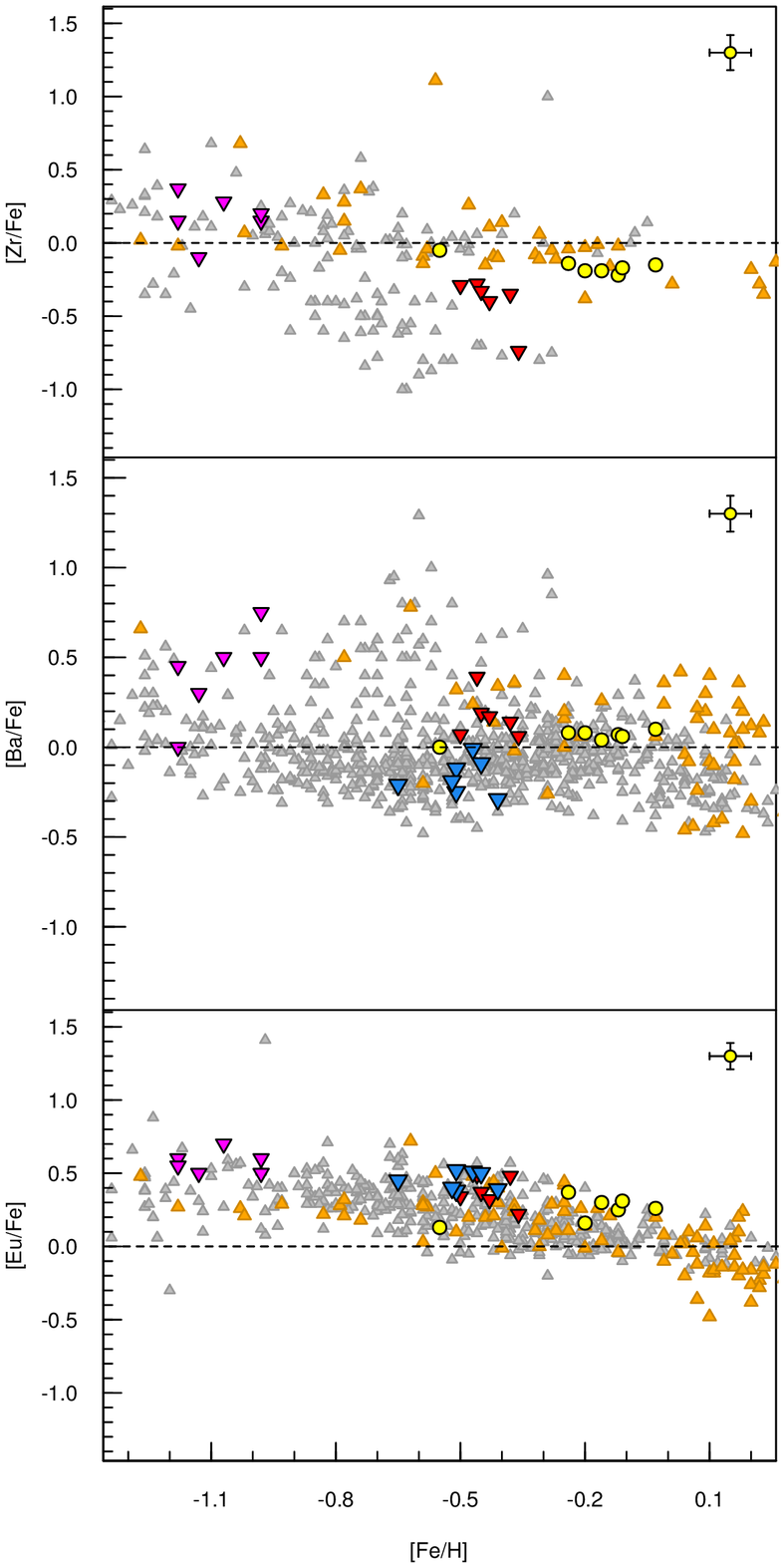}
  \caption{[Zr,Ba, Eu/Fe] vs [Fe/H]. Filled yellow circles are our data for NGC~6528, filled blue triangles: NGC~6440 \citep{munoz17} filled red triangles: NGC 6441 \citep{gratton06},  filled magenta triangles: HP1 \citep{barbuy16}, filled orange triangles: bulge field stars \citep{vanderswaelmen16}, filled  gray triangles: Halo and disk stars \citep{fulbright00,francois07,reddy06,barklem05,venn04}.}

  \label{heavy}
 \end{figure}

We measured three  heavy elements: Zr, Ba and Eu.
We find that the abundance of Ba for NGC~6528 is slightly super-solar ([Ba/Fe]=0.06 dex), showing a similar value to that of NGC~6441 (see Figure \ref{heavy}).  \citet{carretta01} obtained the  Ba abundances in four RHB stars in good agreement with our results (see Figure \ref{comp}).

For [Eu/Fe], the bulge field stars show a decrease with  [Fe/H]. The decrease is due to  the production of iron by SNeIa (Ballero et al. 2007). In the case of Europium, we found a supersolar value ([Eu/Fe]=0.25), in very good agreement with the bulge field star trend and with other bulge GCs including NGC~6440 and NGC~6441 (see figure \ref{heavy}).

\begin{figure} [htb]
\centering
  \includegraphics[width=3.5in,height=3.7in]{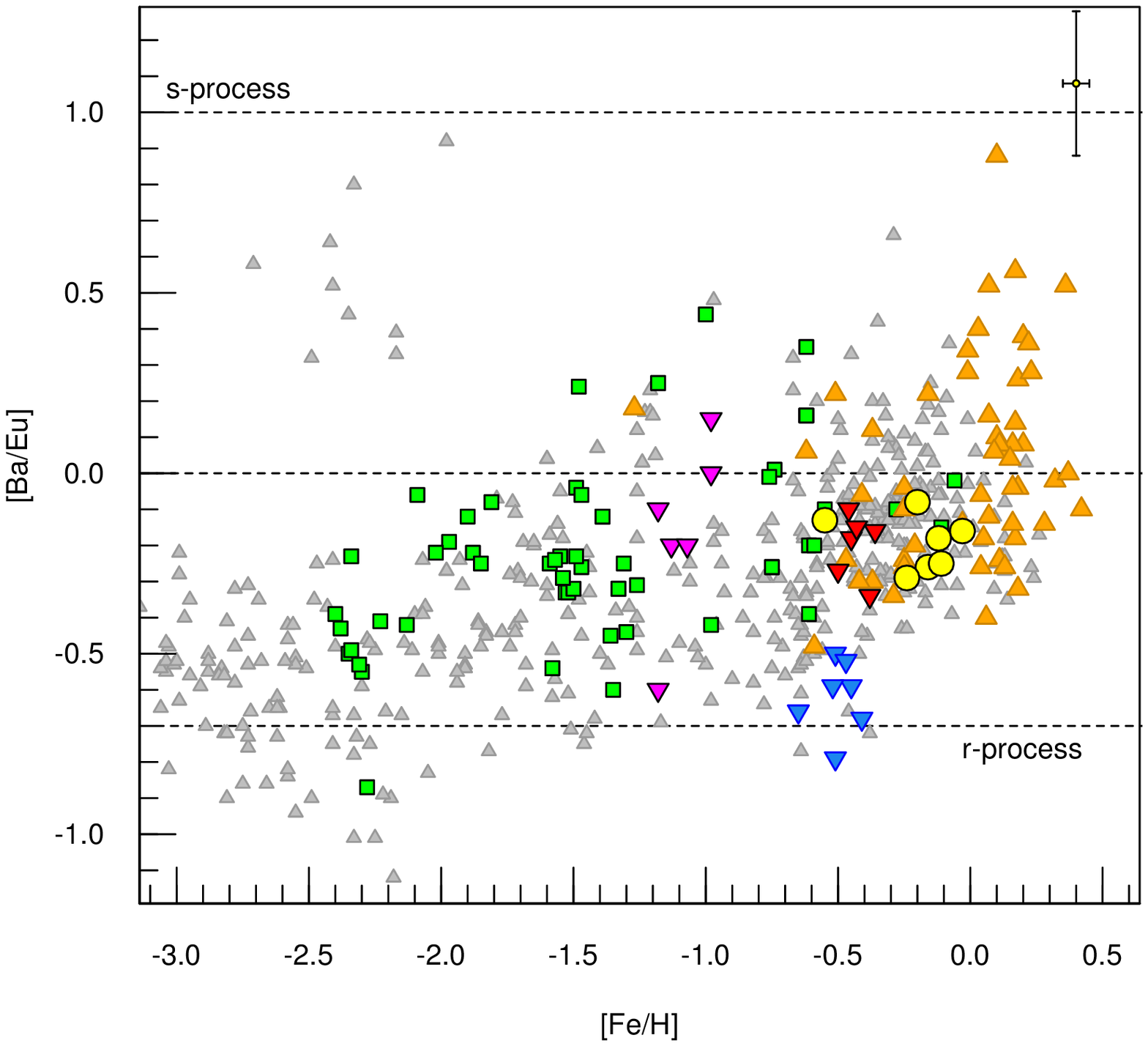}
  \caption{[Ba/Eu] vs [Fe/H]. Filled yellow circles are our data for NGC 6528, filled blue triangles: NGC 6440 \citep{munoz17}  filled red triangles: NGC 6441 \citep{gratton06},  filled magenta triangles: HP1 \citep{barbuy16}, filled orange triangles: bulge field stars \citep{vanderswaelmen16}, filled green square: GCs from  \citet{pritzl05}, filled  grey triangles: Halo and disk stars \citep{fulbright00,francois07,reddy06,barklem05,venn04}.}

  \label{baeu}
 \end{figure}

\begin{figure}[htb]
  \includegraphics[width=3.4in,height=7.9in]{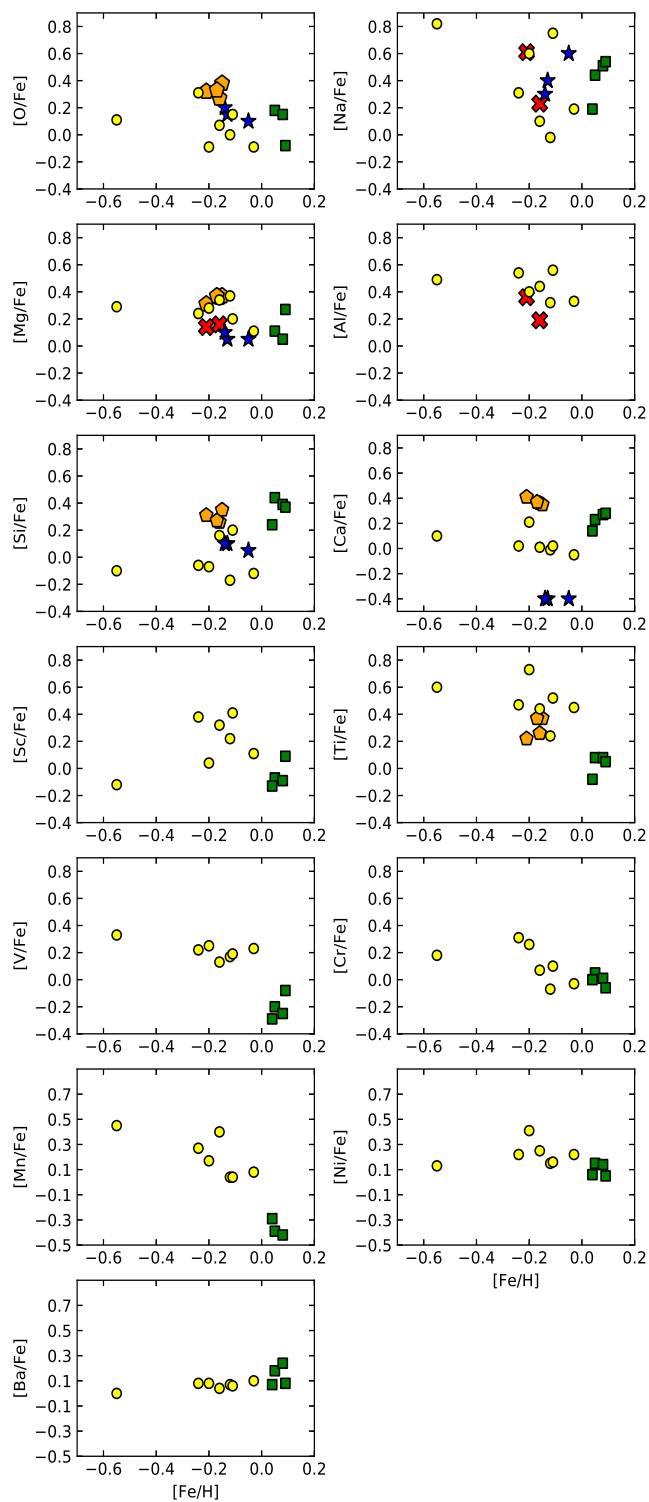}
  \caption{Comparison between this study (yellow filled circles) with other studies for NGC~6528. Green filled squares from \citet{carretta01}, orange filled pentagons from \citet{origlia05}, filled blue stars from \citet{zoccali04}, red filled crosses  from \citet{schiavon17}. }

  \label{comp}
 \end{figure}

In Figure \ref{baeu} we observe [Ba/Eu] versus [Fe/H] which allows us to establish the relative importance between s-process versus r-process. In the case of NGC~6528, we found a value of [Ba/Eu] subsolar,  similar to the values shown by NGC~6441, which suggests that the heavy elements have been produced mainly by explosive events in SNeII. We also noticed very good agreement with the bulge (see figure \ref{baeu}).

Our results are compatible with the results found by \citet{vanderswaelmen16}, which found a turnover in metallicity around  [Fe/H]$\sim$ -0.6 dex, indicating an enrichment from SNe Ia. In this same context, Van der Swaelmen  compared  thick disk stars \citep{bensby04} with a sample from  bulge microlensed stars \citep{bensby13}, he found a small  difference in the chemical enrichment  between bulge and thick disk stars,  apparently associated with a different SN II/SN Ia ratios along time.

The decreasing of Zr with the metallicity for the bulge field stars is evident in figure 11, this trend was detected by \citet{Johnson12}. Our sample of 7 stars of NGC 6528 shows good agreement with the pattern of Zr in this regime of metallicity. Also, we found that our sample    is depleted  in Zr ([Zr/Fe]=-0.16).


%

\section{Conclusions}
This is the second paper of a series of papers about metal rich bulge GCs. In this paper, we have presented a detailed chemical characterization of NGC 6528 from seven red giant members. Using VLT FLAMES/UVES high resolution spectra, we measured 18 chemical elements and we performed a detailed error analysis. This allowed us to make a robust comparison with our first study on NGC 6440 as well as with many other Milky Way objects  (bulge field stars, disk field stars, halo field stars, Galactic GCs).
The most important results of this study are summarized in the following:
\begin{itemize}
\item We found a candidate variable  star (star \#5) in our sample. It turns out to be the most metal poor of the sample ([Fe/H]=-0.55 dex), which may be due to its variaability, but all its chemical patterns show good agreement with the other members of NGC~6528.
\item We obtained  a mean iron abundance of [Fe/H]=-0.14$\pm$0.06 dex  in good agreement with previous literature values.  We did not find a significant spread in iron, although our sample is small (after deleting the one dubious outlier).
\item We measure five iron-peak elements. These show very good agreement with the bulge field stars trend, especially with the bulge GC NGC~6553.
\item The alpha-elements also show good agreement with bulge field stars and NGC~6553.
\item We did not find an extended O-Na anticorrelation. This relationship is more vertical, similar to other Bulge GCs, suggesting that only Na varies and O is fixed.
\item We have found  no Mg-Al anticorrelation. Neither did we  find a real spread in either Al or  in Mg.

\item Heavy elements follow the trend of the bulge field stars and the bulge GCs.
\item $[$Ba/Eu] versus [Fe/H] shows good agreement with NGC 6441. Moreover, it is dominated by r-process material.

\item The origin of NGC 6528  is undoubtedly similar to  the Bulge of the Milky Way, and especially to that of the bulge GC NGC~6553.

   \end{itemize}
   
In all of the chemical comparisons we have made, NGC 6528 exhibits  very good agreement with the general bulge chemical patterns, with the only minor exception being Mn. This behavior,  together with its metallicity, could be indicative of a strong original contamination by SN.

The alpha-elements in the bulge field stars  is a good indicator about the origin, because exhibits an overabundance in these elements \citep{mcwilliam94,zoccali06}. In agreement with high rate of star formation in a early epoch, probably in the first Gyr \citep{lagioia14}. Our discovery for NGC~6528 and for the majority of the bulge GCs analyzed in this study  are compatible with  this behavior.

Furthermore, we note that in the case of the O-Na anticorrelation, several of the Bulge GCs show only a Na variation without a significant O spread (Hp1, NGC 6440, NGC6528, NGC6553), again indicating  a possible similar  origin and evolution  for these bulge GCs, and apparently different from that of Halo GCs. In particular, we highlight that for each chemical pattern in which we compared NGC 6528 with NGC 6553, we found very good agreement, indicating a similar evolution and origin.

According to \citet{dinescu03}   the velocity of NGC~6528 is indication of a radial orbit  confined to the galactic plane.  Also,   its rotation velocity show good agreement with the velocity predicted for the Milky Way bar \citep{dinescu03}.

NGC~6528 is an GC with a age of 11 Gyr \citep{lagioia14} and  one of the most metal rich galactic GC ([Fe/H]=0.14), all the pattern, chemically and dynamically,  support the idea of NGC~6528  is a  bulge GCs with the similar  origin and evolution of the bulge of the Milky Way.

Finally, the most interesting of the above results are the lack of any Na-O or Mg-Al anticorrelation and the homogeneity of the O, Mg and Al abundances, although the sample size is limited. This trend is in general in good agreement with other Bulge GCs (NGC 6441, NGC 6553, NGC 6440). This could indicate a different evolution from that of the typical halo or disk GCs. As usual, additional studies, with much larger samples of clusters as well as stars per cluster, will be needed to clearly define and understand these trends.

\begin{acknowledgements}
This work is Based on observations collected at the European Organisation for Astronomical Research in the Southern Hemisphere under ESO programme ID 093.D-0286.  We gratefully acknowledge use  of data from ESO Public Survey programme   ID 172.B-2002 taken with the VISTA telescope. We gratefully acknowledge support from the Chilean BASAL   Centro de Excelencia en Astrof\'{i}sica
y Tecnolog\'{i}as Afines (CATA).
S.V. gratefully acknowledges the support provided by FONDECYT reg. n. 1170518. C.C.C. is supported by CONICYT (Chile) throught Programa Nacional de Becas de Doctorado 2014 (CONICYT-PCHA/Doctorado Nacional/2014-21141084).D.G. also acknowledges financial support from the Dirección de  Investigación y Desarrollo de la Universidad de La Serena through the Programa de Incentivo a la Investigación de Académicos (PIA-DIDULS). We would also like to thank the referee for his valuable comments and corrections.

\end{acknowledgements}

\bibliographystyle{aa} 
\bibliography{biblio.bib}

\end{document}